\documentclass[english,prl, twocolumn]{revtex4}
\usepackage[latin9]{inputenc}
\usepackage{xcolor}
\usepackage{babel}
\usepackage{amsmath}
\usepackage{amssymb}
\usepackage{graphicx}
\usepackage[unicode=true,pdfusetitle,
 bookmarks=true,bookmarksnumbered=false,bookmarksopen=false,
 breaklinks=false,pdfborder={0 0 1},backref=false,colorlinks=false]
 {hyperref}

\makeatletter
\@ifundefined{textcolor}{}
{%
 \definecolor{BLACK}{gray}{0}
 \definecolor{WHITE}{gray}{1}
 \definecolor{RED}{rgb}{1,0,0}
 \definecolor{GREEN}{rgb}{0,1,0}
 \definecolor{BLUE}{rgb}{0,0,1}
 \definecolor{CYAN}{cmyk}{1,0,0,0}
 \definecolor{MAGENTA}{cmyk}{0,1,0,0}
 \definecolor{YELLOW}{cmyk}{0,0,1,0}
}

\makeatother

\begin{document}
\title{Quantum fluxes at the inner horizon of a spherical charged black hole}
\author{Noa Zilberman}
\email{noazilber@campus.technion.ac.il}

\author{Adam Levi}
\email{leviadam@gmail.com}

\author{Amos Ori}
\email{amos@physics.technion.ac.il}

\affiliation{Department of Physics, Technion, Haifa 32000, Israel}
\date{\today}
\begin{abstract}
In an ongoing effort to explore quantum effects on the interior geometry
of black holes, we explicitly compute the semiclassical flux components
$\left\langle T_{uu}\right\rangle _{ren}$ and $\left\langle T_{vv}\right\rangle _{ren}$
($u$ and $v$ being the standard Eddington coordinates) of the renormalized
stress-energy tensor for a minimally-coupled massless quantum scalar
field, in the vicinity of the inner horizon (IH) of a Reissner-Nordström
black hole. These two flux components seem to dominate the effect
of backreaction in the IH vicinity; and furthermore, their regularization
procedure reveals remarkable simplicity. We consider the Hartle-Hawking
and Unruh quantum states, the latter corresponding to an evaporating
black hole. In both quantum states, we compute $\left\langle T_{uu}\right\rangle _{ren}$
and $\left\langle T_{vv}\right\rangle _{ren}$ in the IH vicinity
for a wide range of $Q/M$ values. We find that both $\left\langle T_{uu}\right\rangle _{ren}$
and $\left\langle T_{vv}\right\rangle _{ren}$ attain finite asymptotic
values at the IH. Depending on $Q/M$, these asymptotic values are
found to be either positive or negative (or vanishing in-between).
Note that having a nonvanishing $\left\langle T_{vv}\right\rangle _{ren}$
at the IH implies the formation of a curvature singularity on its
ingoing section, the Cauchy horizon. Motivated by these findings,
we also take initial steps in the exploration of the backreaction
effect of these semiclassical fluxes on the near-IH geometry.
\end{abstract}
\maketitle

\paragraph*{Introduction.}

The analytically extended Kerr and Reissner-Nordström (RN) metrics,
describing respectively spinning and spherical charged isolated black
holes (BHs), reveal a traversable passage through an inner horizon
(IH) to another external universe \cite{Carter:1966,GravesBrill:1960}.

Consider a traveler intending to access this other universe. To do
so, she must pass through the BH interior, and in particular, through
the IH. What will she encounter along her way? Is her mission doomed
to fail? Does this external universe actually exist? Answering these
questions requires understanding how quantum fields change the internal
geometry of BHs. The most renowned phenomenon in which quantum effects
profoundly transform the classical spacetime picture is the process
of BH evaporation due to Hawking radiation \cite{Hawking:1974,Hawking:1975}.
In fact, already at the classical level, it was demonstrated that
introducing matter (or perturbation) fields on BH backgrounds may
affect their regularity. A notable example is the null weak \cite{Tipler}
curvature singularity that forms along the \emph{Cauchy horizon} (CH,
the IH ingoing section) in both spinning \cite{Ori:1992,Ori:1999,BradyDrozMorsnik:1998,Dafermos:2017}
and spherical charged \cite{Hiscock:1981,PoissonIsrael:1990,OriMassInflation:1991,BradySmith:1995,Piran,Burko:1997,Dafermos:2005}
BHs. The analogous effect of \emph{quantum} perturbations is often
expected to be significantly stronger \cite{BirrellDavies:1978,Hiscock:1980,Ottewill:2000},
but this issue remains inconclusive, making it the main motivation
for this work.

A theoretical framework that lends itself to this problem is the \emph{semiclassical}
formulation of general relativity, considering matter fields as quantum
fields propagating in a classical curved spacetime, obeying the semiclassical
Einstein field equation, given (in units $G=c=1$) by: 
\begin{equation}
G_{\alpha\beta}=8\pi\left\langle T_{\alpha\beta}\right\rangle _{ren}\,.\label{eq:semi}
\end{equation}
Here $G_{\alpha\beta}$ is the Einstein tensor, and the source term
$\left\langle T_{\alpha\beta}\right\rangle _{ren}$ is the renormalized
expectation value of the stress-energy tensor (RSET) associated with
the quantum field. Note the emergent requirement for self-consistency:
spacetime curvature induces a non-trivial stress-energy in the quantum
fields, which in turn deforms the spacetime metric \textemdash{} an
effect known as \emph{backreaction}. A possible way to handle this
complexity is to break the problem into steps of increasing order
in the mutual effect, initially computing $\left\langle T_{\alpha\beta}\right\rangle _{ren}$
for a fixed, classical background metric. But already at this level,
one faces a serious challenge: the computation of the RSET on curved
backgrounds.

Recall that already in flat spacetime the stress-energy tensor of
a quantum field formally diverges, but this is usually handled through
the normal-ordering scheme, which is ill-defined in curved spacetime.
The intricate regularization procedure required in curved spacetime,
along with its inevitable numerical implementation, has made this
computation a decades-lasting hurdle in the study of semiclassical
problems. However, the recently developed \emph{pragmatic mode-sum
regularization} (PMR) method \cite{AAt:2015,AAtheta:2016,AARSET:2016,LeviRSET:2017},
rooted in covariant point-splitting \cite{Christensen:1976,Christensen:1978},
has made this task more accessible. (See, however, earlier works employing
other methods, e.g. \cite{Candelas:1980,Frolov:1982,Fawcett:1983,Candelas-Howard:1984,Howard-Candelas:1984,HowardRSET:1984,Candelas_Jensen:1986,Anderson:1989,Jen_Otte:1989,Anderson:1990,McLau_Jen_Otte:1992,Ander_His_Sam:1995,Otte_Taylor:2011,Taylor:2019}).

The PMR method overcomes the main difficulty in the numerical implementation
of point splitting by treating the coincidence limit analytically,
through construction of ``modewise'' counter-terms. It has been
successfully used in recent years to compute both the vacuum expectation
value $\left\langle \Phi^{2}\right\rangle _{ren}$ and the RSET for
a quantum scalar field $\Phi$ on various BH exteriors \cite{AAt:2015,AAtheta:2016,AARSET:2016,LeviRSET:2017,AAKerr:2017}.
On BH \emph{interiors}, however, only $\left\langle \Phi^{2}\right\rangle _{ren}$
has been computed in that method so far (initially for Schwarzschild
\cite{SchAssaf:2018}, reproducing previous results \cite{Candelas_Jensen:1986}).
Although $\left\langle \Phi^{2}\right\rangle _{ren}$ is not the quantity
most relevant for backreaction, it nevertheless provides valuable
insights for the computation of the more divergent RSET. In particular,
in a recent paper \cite{GroupPhiRN:2019}, $\left\langle \Phi^{2}\right\rangle _{ren}$
was investigated both numerically and analytically inside RN, with
extensive study of the IH vicinity. The RSET trace (for a minimally-coupled
scalar field) was consequently found to \emph{diverge} at the IH.
The following work is a natural continuation of previous ones, providing
novel results for certain key components of the RSET inside a BH \textemdash{}
which directly demonstrate the divergence of semiclassical energy-momentum
fluxes at the CH. \footnote{See also \cite{Taylor:2019}; but note that the unusual quantum state
constructed there does not allow investigating the anticipated semiclassical
CH divergency.}

We hereby consider a spherically-symmetric charged BH, whose geometry
is described by the RN metric:{\small{}
\[
ds^{2}=-f(r)dt^{2}+\frac{1}{f(r)}dr^{2}+r^{2}d\Omega^{2}\,,
\]
}where $d\Omega^{2}=d\theta^{2}+\sin^{2}\theta d\varphi^{2}$, and
$f\left(r\right)\equiv1-2M/r+Q^{2}/r^{2}$ with mass $M$ and charge
$Q$. We consider a non-extremal BH, with $0<Q/M<1$. The event horizon
(EH) and the IH are located at $r=r_{+}$ and $r=r_{-}$ respectively,
with $r_{\pm}\equiv M\pm\sqrt{M^{2}-Q^{2}}$. For later use, we define
the two surface gravity parameters, $\kappa_{\pm}=\left(r_{+}-r_{-}\right)/2r_{\pm}^{2}$.

Upon this background we introduce an (uncharged) minimally-coupled
massless scalar quantum field $\Phi\left(x\right)$, obeying the (covariant)
d'Alembertian equation, $\square\Phi=0$. We decompose the field into
modes, which, owing to the metric symmetries, may be separated into
$e^{-i\omega t}$, spherical harmonics $Y_{lm}\left(\theta,\varphi\right)$,
and a function of $r$ \cite{Group:2018}. The latter is encoded in
the \emph{radial function} $\psi_{\omega l}\left(r\right)$, satisfying:

\begin{equation}
\frac{d^{2}\psi_{\omega l}}{dr_{*}^{2}}+\left[\omega^{2}-V_{l}\left(r\right)\right]\psi_{\omega l}=0\,,\,\,\,\label{eq: Radial_eq}
\end{equation}
with the effective potential
\begin{equation}
V_{l}\left(r\right)=f(r)\left[\frac{l\left(l+1\right)}{r^{2}}+\frac{df/dr}{r}\right]\,.\label{eq: Potential}
\end{equation}
$r_{*}$ is the standard tortoise coordinate defined through $dr/dr_{*}=f(r)$,
varying from $r_{*}\rightarrow-\infty$ at the EH to $r_{*}\to\infty$
at the IH.

In the BH interior $f\left(r\right)<0$, meaning the coordinate $r$
is now timelike. Then, assuming a free incoming wave at the EH, Eq.
(\ref{eq: Radial_eq}) is endowed with the initial condition
\begin{equation}
\psi_{\omega l}\cong e^{-i\omega r_{*}}\,,\,\,\,\,\,\,r_{*}\rightarrow-\infty\,.\label{eq:BC}
\end{equation}

We consider our field in two quantum states: the \emph{Hartle-Hawking}
(HH) state \cite{HH:1976,Israel:1976}, corresponding to a BH in thermal
equilibrium, and the more physically realistic \emph{Unruh} state
\cite{Unruh:1976}, describing an evaporating BH.

We introduce the null Eddington coordinates inside the BH, $u=r_{*}-t$
and $v=r_{*}+t$. The \emph{flux components} of the RSET, $\left\langle T_{uu}\right\rangle _{ren}$
and $\left\langle T_{vv}\right\rangle _{ren}$, are of particular
interest \footnote{Note that the fluxes do not contribute to the RSET trace, which was
shown to diverge in \cite{GroupPhiRN:2019}.}. The reason is threefold. First and foremost, as we shall see, it
is these components that seem to be the most significant for backreaction
near the CH, with a remarkable \emph{accumulating} effect on the form
of the metric (as opposed to minor local distortions associated with
other RSET components). In addition, note that although the classical
RN background contains a non-zero stress-energy tensor (of the sourceless
electromagnetic field), its $T_{uu}$ and $T_{vv}$ components vanish
identically, leaving quantum contributions to prevail. Finally, their
regularization procedure turns out to be especially manageable. Accordingly,
aiming for the ``heart'' of the RSET in the context of backreaction,
this work focuses on the flux components $\left\langle T_{uu}\right\rangle _{ren}$
and $\left\langle T_{vv}\right\rangle _{ren}$ in the IH vicinity.

In the next section we implement the PMR $\theta$-splitting variant
\cite{AAtheta:2016,LeviThetaRSET} to obtain expressions for the renormalized
semiclassical flux components in both quantum states, revealing notable
simplicity when taking the IH limit. We then provide numerical results
for various $Q/M$ values, noting various issues that arise. Finally,
we present a preliminary analysis of backreaction and implications
to the fate of our traveler.

\paragraph*{Developing the near-IH flux expressions.}

In what follows, indices $U$ and $H$ correspond to the Unruh and
HH states, respectively. As mentioned, we shall only consider the
two flux components $\left\langle T_{uu}\right\rangle _{ren}$ and
$\left\langle T_{vv}\right\rangle _{ren}$, and for their uniform
treatment we introduce the symbol $y$, representing either $u$ or
$v$.

The basic PMR expression for the trace-reversed RSET is given in Eq.
(2.6) of Ref. \cite{LeviRSET:2017}. In the case of interest (i.e.
the flux components $\left\langle T_{yy}\right\rangle _{ren}$ evaluated
at $r\to r_{-}$ using $\theta$-splitting), two remarkable simplifications
occur: (i) the PMR counter-term $\tilde{L}_{yy}\left(x,x'\right)$
vanishes \cite{LeviThetaRSET,Sup}; and (ii) since $g_{yy}=0$, $T_{yy}$
coincides with its trace-reversed counterpart. The expression then
simplifies to
\begin{equation}
\left\langle T_{yy}\right\rangle _{ren}\left(x\right)=\frac{1}{2}\lim_{x'\to x}G^{\left(1\right)}\left(x,x'\right)_{,yy'}\,,\label{eq:  point_split}
\end{equation}
where $G^{\left(1\right)}\left(x,x'\right)=\left\langle \left\{ \Phi(x),\Phi(x')\right\} \right\rangle $,
and $\left\{ p(x),q(x')\right\} $ denotes $p(x)q(x')+p(x')q(x)$.
We can also express $G^{\left(1\right)}$ as

\begin{equation}
G^{\left(1\right)}\left(x,x'\right)=\hbar\sum_{l,m}\int_{0}^{\infty}d\omega\,E_{\omega lm}\left(x,x'\right)\,,\label{eq: Hadamard}
\end{equation}
where the mode contributions $E_{\omega lm}\left(x,x'\right)$ inside
a RN BH, in the HH state, are given by
\begin{align*}
E_{\omega lm}^{H}\left(x,x'\right) & =\coth\tilde{\omega}\left[J^{R}+J^{L}+\left(\cosh\tilde{\omega}\right)^{-1}J^{RL}\right]
\end{align*}
(cf. Eq. (4.3) in \cite{Group:2018}) where 
\[
J^{R}=\left\{ f_{\omega lm}^{R}\left(x\right),f_{\omega lm}^{R*}\left(x'\right)\right\} ,\,J^{L}=\left\{ f_{\omega lm}^{L}\left(x\right),f_{\omega lm}^{L*}\left(x'\right)\right\} 
\]
 and 
\[
J^{RL}=2\Re\left[\rho_{\omega l}^{up}\left\{ f_{\omega lm}^{R}\left(x\right),f_{(-\omega)lm}^{L*}\left(x'\right)\right\} \right]\,.
\]
Here $\tilde{\omega}\equiv\pi\omega/\kappa_{+}$, the star denotes
complex conjugation, and $\Re$ marks the real part. Hereafter, $\rho_{\omega l}^{up}\,\left(\tau_{\omega l}^{up}\right)$
represents the reflection (transmission) coefficient for the ``up''
modes outside the BH \cite{Group:2018}. The mode functions $f_{\omega lm}^{R,L}\left(x\right)$
are given by 
\[
f_{\omega lm}^{R,L}(x)=\frac{1}{r\sqrt{4\pi\left|\omega\right|}}Y_{lm}(\theta,\varphi)\tilde{f}_{\omega l}^{R,L}\left(t,r\right)
\]
where $\tilde{f}_{\omega l}^{R}=e^{-i\omega t}\psi_{\omega l}(r)$
and $\tilde{f}_{\omega l}^{L}=e^{i\omega t}\psi_{\omega l}(r)$, and
$\psi_{\omega l}\left(r\right)$ is the aforementioned radial function
solving Eq. (\ref{eq: Radial_eq}) with the initial condition (\ref{eq:BC}).
(For more details see \cite{Group:2018}.)

A similar expression exists for the Unruh-state counterpart, $E_{\omega lm}^{U}$.
In what follows, we shall describe the analysis for the HH state solely.
For the Unruh state the analysis is similar and we shall merely quote
final results below (with the more detailed derivation deferred to
\cite{Sup}). Note that due to time-inversion symmetry of the HH state
(unlike the Unruh state), $\left\langle T_{uu}\right\rangle _{ren}^{H}=\left\langle T_{vv}\right\rangle _{ren}^{H}$
everywhere.

We are interested in the asymptotic behavior at the IH, where the
effective potential $V_{l}\left(r\right)$ vanishes like $f\propto r-r_{-}$.
Hence the radial equation (\ref{eq: Radial_eq}) for $\psi_{\omega l}$
admits the general asymptotic solution $A_{\omega l}e^{i\omega r_{*}}+B_{\omega l}e^{-i\omega r_{*}}$
(with constant coefficients $A_{\omega l},\,B_{\omega l}$ ), which
in turn implies
\begin{equation}
\tilde{f}_{\omega l}^{R}\cong A_{\omega l}e^{i\omega u}+B_{\omega l}e^{-i\omega v},\,\,\tilde{f}_{\omega l}^{L}\cong A_{\omega l}e^{i\omega v}+B_{\omega l}e^{-i\omega u}\,.\label{eq: Free_f}
\end{equation}

Equations (\ref{eq:  point_split},\ref{eq: Hadamard}) yield
\[
\left\langle T_{yy}\right\rangle _{ren}^{H}\left(x\right)=\frac{\hbar}{2}\lim_{x'\to x}\sum_{l,m}\int_{0}^{\infty}d\omega\,E_{\omega lm}^{H}\left(x,x'\right)_{,yy'}\,.
\]
It is interesting to inspect $E_{\omega lm}^{H}\left(x,x'\right)_{,yy'}$
within the near-IH approximation (\ref{eq: Free_f}). Consider, for
example, the contribution coming from the $J^{R}$ term. Focusing
for concreteness on $y=u$, we readily see that the $\partial_{uu'}$
operator annihilates the terms depending on $v$ in Eq. (\ref{eq: Free_f}).
Also, $r_{,u}=f/2\propto r-r_{-}$ vanishes at $r\to r_{-}$, altogether
yielding at the limit $\left(u',v',\varphi'\right)\to\left(u,v,\varphi\right)$
(corresponding to $\theta$-splitting) and $r\to r_{-}$:
\begin{equation}
J_{,uu'}^{R}\to\left\{ Y_{lm}(\theta,\varphi),Y_{lm}^{*}(\theta',\varphi)\right\} \left|A_{\omega l}\right|^{2}\,.\label{eq:J^R_uu'}
\end{equation}
Remarkably, although $J^{R}$ itself does contain terms like $\propto e^{i\omega(v+u)}=e^{2i\omega r_{*}}$
at the IH limit, $J_{,uu'}^{R}$ is free of such oscillatory terms
\textemdash{} and is in fact entirely independent of $r_{*}$ (and
$t$). This simplification occurs for all three ``$J$'' terms in
the expression for $E_{\omega lm}^{H}\left(x,x'\right)_{,uu'}$. Combining
their contributions and summing over $m$, one readily obtains at
the IH
\begin{align}
\left\langle T_{uu}\right\rangle _{ren}^{H} & =\hbar\lim_{\delta\theta\to0}\sum_{l=0}^{\infty}\frac{2l+1}{8\pi}P_{l}\left(\cos\delta\theta\right)F_{l}^{H}\,,\label{eq: Final_a_H}
\end{align}
where $\delta\theta\equiv\theta'-\theta$, and $F_{l}^{H}\equiv\int_{0}^{\infty}d\omega\,\hat{E}_{\omega l}^{H}$
where
\begin{equation}
\hat{E}_{\omega l}^{H}=\frac{\omega\coth\tilde{\omega}}{\pi r_{-}^{2}}\left[\left|A_{\omega l}\right|^{2}+\text{cosh}^{-1}\tilde{\omega}\,\Re\left(\rho_{\omega l}^{up}A_{\omega l}B_{\omega l}\right)\right]\label{eq:integrand}
\end{equation}
(see fuller derivation in \cite{Sup}).

The sequence $F_{l}^{H}$ appearing in Eq. (\ref{eq: Final_a_H})
approaches a non-vanishing constant $\beta\equiv F_{l\to\infty}^{H}$.
One can show \cite{Sup}, analytically, that $\beta=\left(\kappa_{-}^{2}-\kappa_{+}^{2}\right)/24\pi r_{-}^{2}$.
Taking the $\delta\theta\to0$ limit (using the methods of Ref. \cite{AAtheta:2016};
see also \cite{Sup}), we obtain the final result
\begin{align}
\left\langle T_{uu}^{-}\right\rangle _{ren}^{H}=\left\langle T_{vv}^{-}\right\rangle _{ren}^{H} & =\hbar\sum_{l=0}^{\infty}\frac{2l+1}{8\pi}\left(F_{l}^{H}-\beta\right)\,.\label{eq: Final_H}
\end{align}
Here, the upper $"-"$ index indicates the IH limit.

The analogous Unruh-state expression is \cite{Sup}:
\begin{align}
\left\langle T_{yy}^{-}\right\rangle _{ren}^{U} & =\left\langle T_{yy}^{-}\right\rangle _{ren}^{H}+\hbar\sum_{l=0}^{\infty}\frac{2l+1}{8\pi}\Delta F_{l\left(yy\right)}^{U}\,,\label{eq: Final_U}
\end{align}
where $\Delta F_{l\left(yy\right)}^{U}\equiv\int_{0}^{\infty}d\omega\,\Delta\hat{E}_{\omega l\left(yy\right)}^{U}$
and
\begin{equation}
\Delta\hat{E}_{\omega l\left(yy\right)}^{U}=\frac{\omega}{2\pi r_{-}^{2}}\left(1-\coth\tilde{\omega}\right)\left|\tau_{\omega l}^{up}\right|^{2}\left(\left|A_{\omega l}\right|^{2}+\delta_{y}^{v}\right)\,.\label{eq:  DE_U}
\end{equation}

Note that the two Unruh-state flux components are not independent:
From energy-momentum conservation, $4\pi r^{2}\left(\left\langle T_{uu}(x)\right\rangle _{ren}^{U}-\left\langle T_{vv}(x)\right\rangle _{ren}^{U}\right)$
is \emph{constant} (it is actually the Hawking outflux; see \cite{Sup}).

\paragraph*{Numerical results.}

Recalling the Wronskian relation $\left|\tau_{\omega l}^{up}\right|^{2}=1-\left|\rho_{\omega l}^{up}\right|^{2}$,
the final expressions (\ref{eq: Final_H},\ref{eq: Final_U}) for
the near-IH fluxes in both quantum states reveal simple dependence
on $A_{\omega l},\,B_{\omega l}$ and $\rho_{\omega l}^{up}$. We
numerically compute $A_{\omega l}$ and $B_{\omega l}$ by integrating
the radial equation (\ref{eq: Radial_eq}) from $r_{+}$ to $r_{-}$
(and $\rho_{\omega l}^{up}$ likewise, by solving the radial equation
outside the BH). We then compute the three flux quantities $\left\langle T_{yy}^{-}\right\rangle _{ren}$
(that is $\left\langle T_{yy}^{-}\right\rangle _{ren}^{H},\,\left\langle T_{uu}^{-}\right\rangle _{ren}^{U}$
and $\left\langle T_{vv}^{-}\right\rangle _{ren}^{U}$) at the IH,
as prescribed in Eqs. (\ref{eq: Final_H},\ref{eq: Final_U}). For
further numerical details, see \cite{Sup}. We find exponential convergence
of both the integral over $\omega$ (entailed in $F_{l}^{H},\Delta F_{l}^{U}$)
and the sum over $l$, for all three quantities $\left\langle T_{yy}^{-}\right\rangle _{ren}$,
as they attain well-defined \emph{finite} values. Note that a finite
non-vanishing $\left\langle T_{vv}^{-}\right\rangle _{ren}$ implies
a \emph{curvature singularity} at the CH, since transforming to a
regular Kruskal-like coordinate $V=-e^{-\kappa_{-}v}$ yields $\left\langle T_{VV}^{-}\right\rangle _{ren}\propto e^{2\kappa_{-}v}\to\infty$.

\begin{figure}[h!]
\centering \includegraphics[width=8cm]{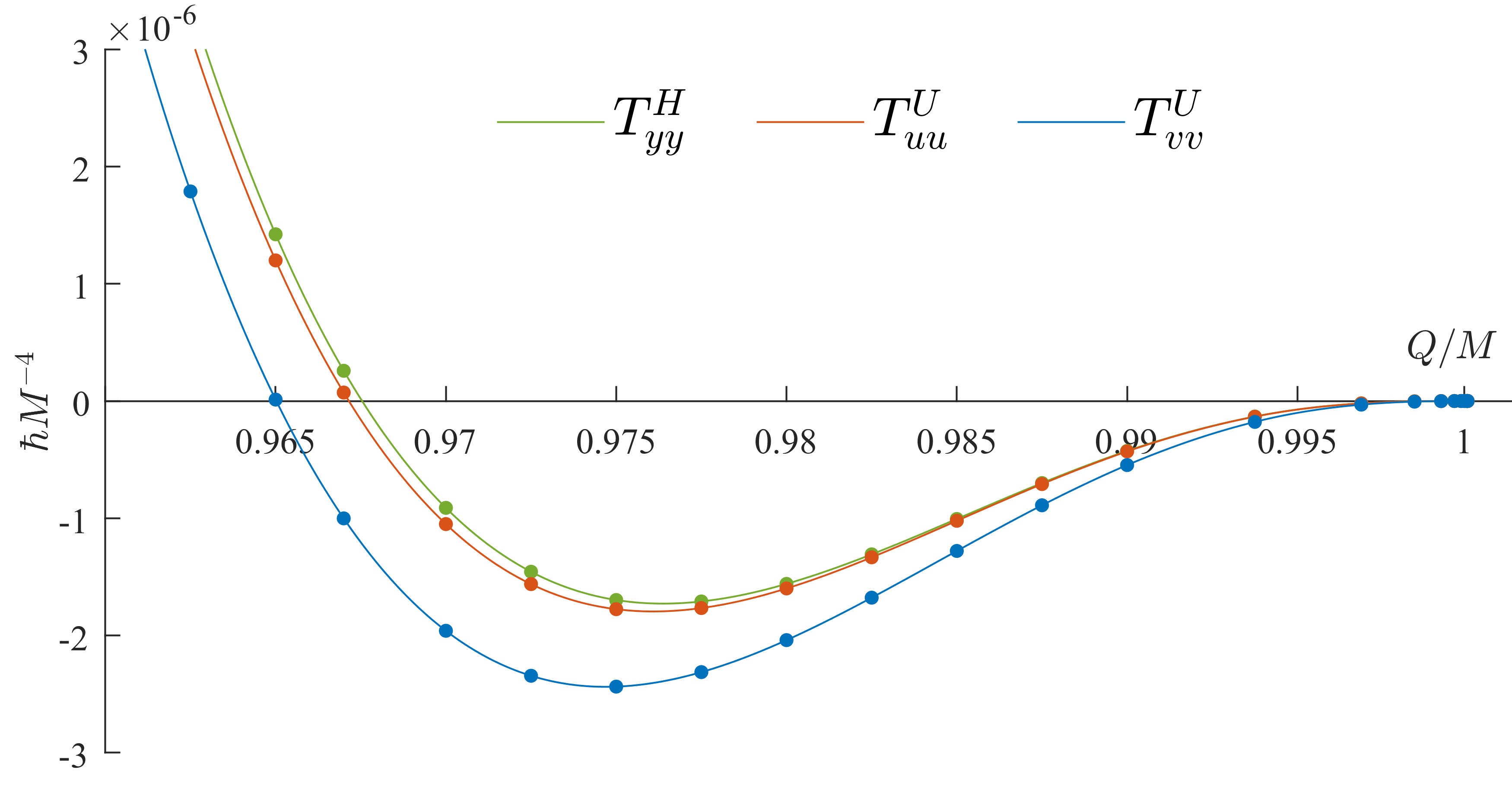}\caption{$\left\langle T_{yy}^{-}\right\rangle _{ren}$ (namely $\left\langle T_{uu}^{-}\right\rangle _{ren}^{U},\,\left\langle T_{vv}^{-}\right\rangle _{ren}^{U}$,
and \textcolor{black}{${\color{purple}{\normalcolor \left\langle T_{uu}^{-}\right\rangle _{ren}^{H}=\left\langle T_{vv}^{-}\right\rangle _{ren}^{H}}}$})
as a function of $Q/M$. The points correspond to the numerical data,
while the solid curve is interpolated.}
\label{transition}
\end{figure}

Remarkably, the three quantities $\left\langle T_{yy}^{-}\right\rangle _{ren}$
may be either positive or negative, depending on $Q/M$. We find that
sufficiently close to extremality all three flux components become
negative, whereas further away from extremality they are all positive.
Whether the diverging $\left\langle T_{VV}^{-}\right\rangle $ is
positive or negative is crucial for the nature of tidal deformation
(contraction vs. expansion), a point expanded hereafter. Figure \ref{transition}
displays the three flux quantities $\left\langle T_{yy}^{-}\right\rangle _{ren}$
in the range $0.96<Q/M<1$, exhibiting the transition from positive
to negative values at around $Q/M\sim0.97$. More precisely, the sign
change occurs at $Q/M$ values of $q_{v}^{U}\cong0.9650,\,q_{u}^{U}\cong0.9671$
and $q_{y}^{H}\cong0.9675$ for $\left\langle T_{vv}^{-}\right\rangle _{ren}^{U},\,\left\langle T_{uu}^{-}\right\rangle _{ren}^{U}$
and $\left\langle T_{yy}^{-}\right\rangle _{ren}^{H}$, respectively.

\begin{figure}[h!]
\centering \includegraphics[width=8cm]{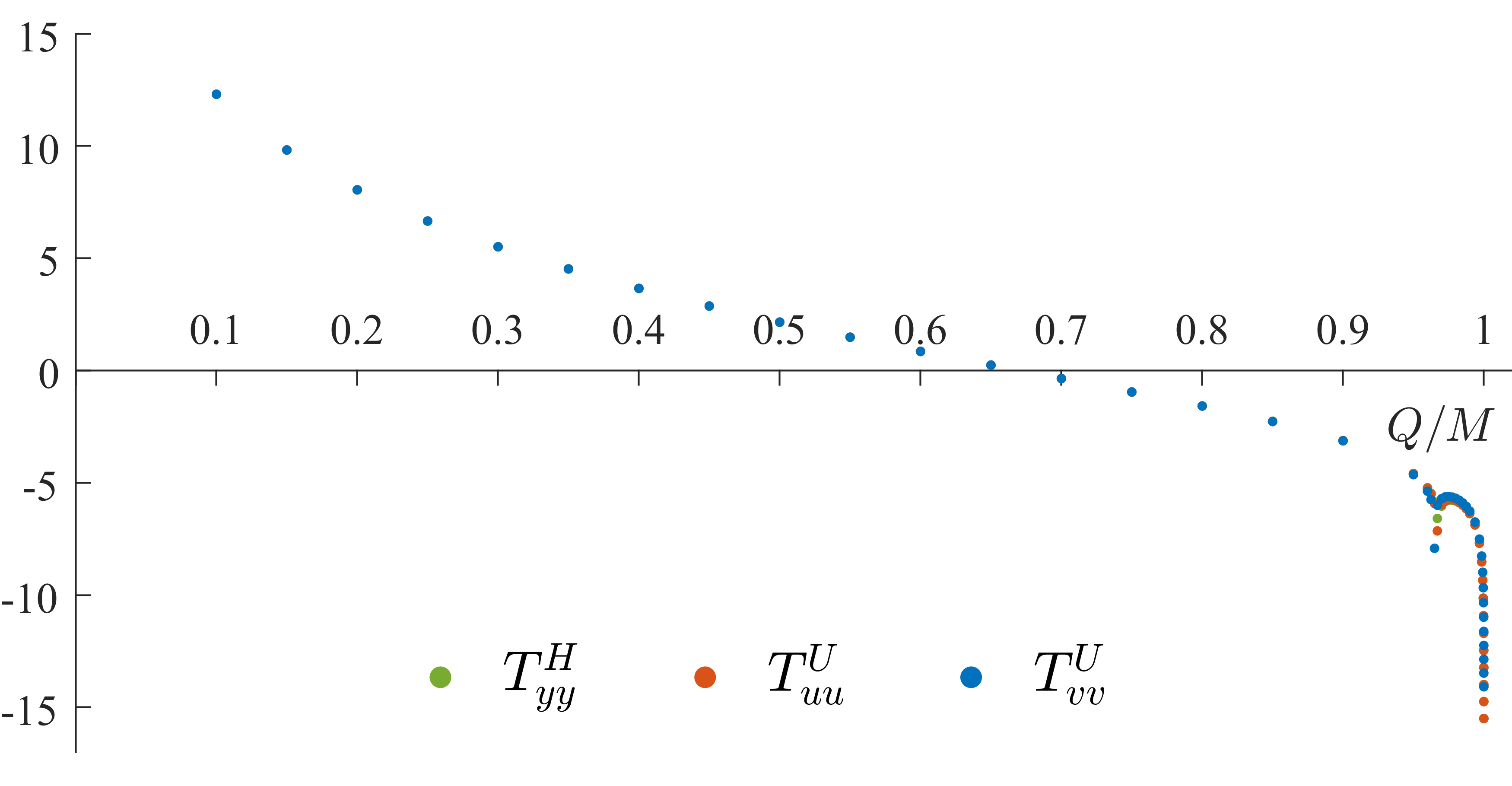}\caption{$\log_{10}\left|\left\langle T_{yy}^{-}\right\rangle _{ren}/\hbar M^{-4}\right|$
for a wider $Q/M$ range. The steep drop at $\sim0.97$ corresponds
to the fluxes changing sign. Note that in most $Q/M$ values the three
quantities are indistinguishable here.}
\label{generalregion}
\end{figure}

Figure \ref{generalregion} displays the three flux quantities in
a wider range $0.1\leq Q/M<1$. Note the very rapid increase in the
fluxes as $Q/M$ decreases. This is perhaps not too surprising, since
a decrease in $Q/M$ implies an (even faster) decrease in $r_{-}/M$,
and correspondingly an increasing curvature at the IH.

Another notable feature is the decay of the fluxes as $Q/M\to1$.
Remarkably, in the near-extremal domain (characterized by $\left|Q/M-1\right|\ll1$),
the flux computation lends itself to analytical treatment (which we
defer to a future paper \cite{Near-extremality}), leading to excellent
agreement with the numerical data illustrated on the rightmost part
of Fig. \ref{transition}.

\paragraph*{Backreaction near the CH.}

The semiclassical backreaction, being of order $\propto\hbar/M^{2}=(m_{p}/M)^{2}$
(where $m_{p}$ denotes the Planck mass), is basically an extremely
weak effect for macroscopic BHs. For instance, for astrophysical BHs
it is typically $<10^{-75}$. However, these effects accumulate along
the EH, causing its area to drastically shrink upon evaporation. Likewise,
as we shall shortly see, semiclassical effects may also accumulate
near the CH (and in addition, they become singular there). Thus, semiclassical
backreaction is presumably negligible \textemdash{} and hence the
actual backreacted geometry should be well approximated by the original
RN metric \textemdash{} as long as (i) the BH hasn't had the chance
yet to significantly evaporate (that is, the $v$ interval since the
BH formation is much smaller than the evaporation timescale), and
(ii) we are not too close to the CH.

To address backreaction, we write the general spherically-symmetric
metric in double-null coordinates as $-e^{\sigma}dudv+r^{2}d\Omega^{2}$.
The two unknown metric functions, $r\left(u,v\right)$ and $\sigma\left(u,v\right)$,
are to be determined from the semiclassical Einstein equation (\ref{eq:semi}).
This system contains \emph{constraint equations}, which are two independent
ODEs (one along each null direction) that involve the flux components
$\left\langle T_{yy}^{-}\right\rangle _{ren}$ only; and \emph{evolution
equations}, which are two coupled PDEs involving $\left\langle T_{uv}\right\rangle _{ren}$
and $\left\langle T_{\theta\theta}\right\rangle _{ren}$. Our analysis
will mainly rely on the two constraint equations, which we write uniformly
as
\begin{align}
r_{,yy} & -r_{,y}\sigma_{,y}=-4\pi r\left\langle T_{yy}\right\rangle _{ren}\,.\label{eq:constraints}
\end{align}

To proceed, we shall now restrict the analysis to the \emph{weak-backreaction}
domain, in which $r,\,\sigma_{,y}$ and $\left\langle T_{yy}\right\rangle _{ren}$
(but not necessarily $r_{,y}$) are still well approximated by their
original RN background values. \footnote{By its very definition, this weak-backreaction domain must satisfy
restrictions (i,ii) mentioned above.} Correspondingly, in what follows we  consider the RN-background RSET
and explore its backreaction effect via the semiclassical Einstein
equation.

Furthermore, we shall focus on the \emph{near-CH} portion of this
weak-backreaction domain \footnote{This is the region in which $e^{-\kappa_{-}(u+v)}$ is already $\ll1$
(hence in the RN background $r,\,\sigma_{,y}$ and $\bigl\langle T_{yy}\bigr\rangle_{ren}$
are well approximated by their near-CH values); and correspondingly,
the drift effect in $r$ is already present \textemdash{} but still
hasn't accumulated much.}. In this region, we may replace the right hand side of Eq. (\ref{eq:constraints})
by the constant $-4\pi r_{-}\left\langle T_{yy}^{-}\right\rangle _{ren}$,
and $\sigma_{,y}$ by $-\kappa_{-}$ (its near-CH value in RN). We
obtain a trivial linear ODE for $r_{,y}$, which is easily solved.
After an exponentially decaying term ($\propto e^{\sigma}$) is dropped,
we are left with 
\begin{align}
r_{,y} & \cong-4\pi(r_{-}/\kappa_{-})\left\langle T_{yy}^{-}\right\rangle _{ren}\,.\label{eq:r-1st-derivative}
\end{align}

This result expresses a small but steady asymptotic drift of $r\left(u,v\right)$
in both null directions. In the long run (i.e. at sufficiently large
$u$ and/or $v$) this drift would result in a major deviation of
$r$ from its RN value \textemdash{} which would eventually lead us
away from the weak-backreaction domain.

From Eq. (\ref{eq:r-1st-derivative}) it becomes clear that this remarkable
accumulative effect is dictated solely by the flux components, namely,
it is independent of the other RSET components.

To discuss the physical implications of this result, let us assume
our infalling traveler moves towards the IH ingoing section, and approaches
the near-IH domain where the semiclassical drift is present. We shall
consider now the effect of the drift in the $v$ direction. \footnote{Both $\left\langle T_{uu}^{-}\right\rangle _{ren}$ and $\left\langle T_{vv}^{-}\right\rangle _{ren}$
are associated with a drift effect at the CH vicinity, but as mentioned,
only $\left\langle T_{vv}^{-}\right\rangle _{ren}$ induces a singular
effect there. The effect of $\left\langle T_{uu}^{-}\right\rangle _{ren}$
may be associated with a steady drift of $r$ \emph{along the CH.}} We emphasize that although the near-CH drift in $r$ is very ``slow''
in terms of $v$ (i.e. $r_{,v}\ll1$), it actually happens at an exceedingly
fast rate for our infalling traveler \textemdash{} which (in the fiducial
RN geometry) would arrive the CH at a \emph{finite} proper time \footnote{In particular, recall that $dr/d\tau\propto dv/d\tau\propto e^{\kappa_{-}v}$.}.
The nature of this physical effect may crucially depend on the sign
of $\left\langle T_{vv}^{-}\right\rangle _{ren}$ \textemdash{} and
hence on the value of $Q/M$. For $Q/M<q_{v}^{U,H}$, $\left\langle T_{vv}^{-}\right\rangle _{ren}>0$
and correspondingly our traveler will undergo sudden \emph{contraction}.
However, for $Q/M>q_{v}^{U,H}$, $\left\langle T_{vv}^{-}\right\rangle _{ren}$
is negative \textemdash{} implying an abrupt \emph{expansion.}

This analysis still needs to be extended to the domain of strong backreaction,
which actually entails two types of extensions: (i) to the domain
of very late time (i.e. very large $v$), in which significant evaporation
has already occurred \footnote{Obviously this extension is needed in Unruh state only.},
and (ii) to the region very close to the CH.

\paragraph*{Discussion.}

Motivated by long-standing expectations that semiclassical effects
may drastically influence the interior geometry of spinning or charged
BHs, this work focused on the RSET flux components (for a minimally-coupled
massless scalar field), in the IH vicinity, on a fixed RN background.
We presented novel results for the flux components in the Unruh and
HH states for various $Q/M$ values. Both flux components $\left\langle T_{uu}\right\rangle _{ren}$
and $\left\langle T_{vv}\right\rangle _{ren}$ \textemdash{} in both
quantum states \textemdash{} exhibit \emph{finite} asymptotic values
at the IH. Recall, however, that a non-vanishing finite $\left\langle T_{vv}\right\rangle _{ren}$
implies unbounded curvature (and unbounded tidal force) at the CH
($v\to\infty$), because the corresponding Kruskal-like component
$\left\langle T_{VV}\right\rangle _{ren}$ then diverges as $e^{2\kappa_{-}v}$.

Hiscock \cite{Hiscock:1980} previously demonstrated that in the Unruh
state in a Kerr-Newman BH either $\left\langle T_{uu}^{-}\right\rangle _{ren}$
or $\left\langle T_{vv}^{-}\right\rangle _{ren}$ (or possibly both)
are non-vanishing \textemdash{} indicating that the corresponding
Kruskal fluxes diverge on at least \emph{one of the two IH sections}.
Still, this result left the semiclassical CH singularity inconclusive:
Note that it is exclusively the \emph{ingoing} section of the IH which
maintains the causal and physical role of a CH in an astrophysical
BH. \footnote{In particular, a semiclassical divergence that occurs at the \emph{outgoing}
IH section of the Kerr-Newman background is not expected to realize
in a realistic BH produced by gravitational collapse.} Our results show that both $\left\langle T_{uu}^{-}\right\rangle _{ren}$
\emph{and} $\left\langle T_{vv}^{-}\right\rangle _{ren}$ are generically
nonvanishing \textemdash{} demonstrating for the first time the divergence
of the Kruskal flux component $\left\langle T_{VV}\right\rangle _{ren}\propto e^{2\kappa_{-}v}$
at the CH.

It is also worth comparing the semiclassical RSET divergence $\propto e^{2\kappa_{-}v}$
found here with its classical counterpart. Classical perturbations
are known to give rise to curvature divergence at the CH, typically
like $v^{-n}e^{2\kappa_{-}v}$ (with $n$ a positive integer depending
on the type of perturbation) \cite{Hiscock:1981,OriMassInflation:1991,OriLinear:1999}.
In this sense, the aforementioned semiclassical divergence at the
CH is stronger than the one associated with classical perturbations.

Our numerical results indicate that all flux components change their
signs at around $Q/M\sim0.97$, being negative for larger $Q/M$ and
positive (and typically much larger) for smaller $Q/M$ values. The
sign may have crucial implications to the nature of the tidal effect:
catastrophic contraction (for $\left\langle T_{vv}^{-}\right\rangle _{ren}>0$)
vs. expansion (for $\left\langle T_{vv}^{-}\right\rangle _{ren}<0$).

We also made initial steps towards analyzing the semiclassical backreaction
effects of the fluxes on the near-CH geometry (in both the Unruh and
HH states). The result expressed in Eq. (\ref{eq:r-1st-derivative})
hints for drastic deformation of the area coordinate $r$ on approaching
the CH. However, the analysis provided here was rather preliminary.
It should be extended, as mentioned, beyond the domain of weak backreaction.
In particular, this picture may change in the next iteration, in which
the RSET is re-evaluated with respect to the \emph{backreacted }geometry.

Other obvious extensions are in order. First, it would be worthwhile
to generalize the analysis to all RSET components, and also to the
entire interior domain $r_{-}<r<r_{+}$. More importantly, this investigation
should be extended from the scalar to the more realistic electromagnetic
quantum field \textemdash{} and in addition, from the spherical RN
background to the astrophysically much more relevant background of
a spinning BH.
\begin{acknowledgments}
This work was supported by the Asher Fund for Space Research at the
Technion, and by the Israel Science Foundation under Grant No. 600/18.
\end{acknowledgments}

\end{document}